\documentclass[cpl,twocolumn]{revtex4}%
\usepackage{amsfonts}
\usepackage{amsmath}
\usepackage{amssymb}
\usepackage{graphicx}%
\begin{document}
\title{An optical analogue for rotating BTZ black holes}
\author{Ling Chen}
\affiliation{School of Mathematics and Physics, China University of Geosciences, Wuhan
430074, China}
\author{Hongbin Zhang}
\affiliation{School of Mathematics and Physics, China University of Geosciences, Wuhan
430074, China}
\author{Baocheng Zhang}
\email{zhangbaocheng@cug.edu.cn}
\affiliation{School of Mathematics and Physics, China University of Geosciences, Wuhan
430074, China}
\keywords{BTZ black holes, optical vortex, inner horizon, analogue gravity}
\begin{abstract}
We demonstrate an optical realization for the rotating BTZ black hole using
the recent popular photon fluid model in an optical vortex but with a new
proposed expression for the optical phase. We also give the numerical
realization for the optical vortex to ensure that it can be generated
experimentally. Different from the earlier suggestions for the analogue
rotating black holes, our proposal includes an inner horizon in the analogue
black hole structure. Such structure can keep for a long distance for the
convenience of observing analogue Hawking or Penrose radiations.

\end{abstract}

\pacs{04.20.-q, 04.60.-m, 04.80.-y}
\maketitle

\section{Introduction}

The analogue gravity \cite{blv11} had been well-known to provide a prospective
avenue to study the properties of black holes. This stemmed from the thought
put forward initially by Unruh \cite{wgu81}, in which it suggested the
kinematical analogue between the motion of sound wave in a convergent fluid
flow and the motion of a scalar field in the background of Schwarzschild
spacetime. For the analogue, the inhomogeneous flowing fluid was considered as
the spacetime background and the sound waves as a massless scalar field could
move under this spacetime background, but the sound waves cannot propagate
through a surface which is determined by the relation that the fluid velocity
is equal to the local sound velocity and behaves like the event horizon of a
gravitational black hole. Thus, the analogue fluid can present the many
interesting phenomena about the black holes, such as the event horizon
\cite{pkl08}, Hawking radiations \cite{ngks19}, and so on. Due to this, many
other physical systems, such as Bose-Einstein condensates (BEC) \cite{gaz00},
superfluid He \cite{jv98}, slow light \cite{us03}, electromagnetic waveguide
\cite{su05}, light in a nonlinear liquid \cite{efb12}, laser pulse filaments
\cite{bcc10}, trapped-ion systems \cite{ylc20}, i.e. see the review
\cite{blv11}, have been suggested as the physical systems of the analogue
gravity. However, most of these analogue systems were made for the (1+1)
dimensional spacetime, and the analogue phenomena about Schwarzschild black
holes were usually presented.

Actually, there are also many proposals to simulate other types of black holes
in the perspective of analogue gravity, such as the rotating black holes
\cite{vw05,mco09,gl17,obf18,tt20} which could be relevant to the astrophysics,
AdS and dS black holes \cite{dlt16,sh16} for understanding the AdS/CFT or
gravity/fluid duality. But, it is not easy to find the corresponding physical
systems to finish the analogue with the feasible velocity distribution. A
recent experiment \cite{vmf18} used the photon fluid \cite{cc13} to realized
the analogue for the rotating black hole. The photon fluid can carry the
angular momentum and are the ideal candidate for this kind of analogue. In
their experiment, the rotating analogue black hole is presented by an optical
vortex propagating in a self-defocusing medium at room temperature
\cite{vmf18,fm08,tt17}, in\ which an ergosphere is formed to enclose the
internal region in which rotating energy could be extracted \cite{bfw20}. In
this paper, we will suggest a new type of optical vortex to simulate a new
structure about BTZ (Ban\~{a}do, Teitelboim, Zanelli) \cite{btz92} black hole
spacetime, since the earlier models cannot include the inner horizon and
present only the equatorial slice of the Kerr geometry.

As well-known, the BTZ black hole is a solution to Einstein field equation in
a (2+1) dimensional theory of gravity with a negative cosmological constant.
The (2+1) dimensional theory of gravity is usually considered as a useful
model to explore the foundations of classical and quantum gravity
\cite{sc951,tz13,zbc13,lgz17}, and its solution of BTZ black holes is
considered as the effective avenue to realize this to some extent
\cite{sc952}. The BTZ black hole has the common features as a general black
hole, i.e. it has an event horizon and an inner horizon in rotating case, it
appears as the final state of collapsing matter, and it has thermodynamic
properties much like those of a (3+1)-dimensional black hole. Meanwhile, the
BTZ black holes has some important different properties from the Schwarzschild
or Kerr black hole, i.e. it is asymptotically anti-de Sitter rather than
asymptotically flat, and has no curvature singularity at the origin, which
provide a simple platform for understanding the AdS/CFT correspondence or the
properties of black holes relevant to string theory \cite{sc98}. Therefore, it
is significant to find the analogue systems to present the structure of BTZ
black holes in the laboratory. It is noted that the nonrelativistic acoustic
metric can match the rotating BTZ black hole metric in form \cite{gl17}, but
the matching form for the fluid velocity is hard to be realized in the actual
experimental systems. In the paper, we will give a method to simulate the
metric approximately but realize nearly all the structures required by the BTZ
black hole spacetime.

The paper is organized as follows. In the second section, the optical analog
of BTZ black holes is reviewed. The corresponding quantities between the BTZ
black hole and the acoustic metric are given, and the formation conditions for
the horizon are depicted with a figure. In the third section, we discuss a
concrete model using the optical vortex to realize the spacetime structure of
BTZ black holes. In particular, we use the method of numerical simulation to
present the feasibility of our suggested optical vortex. Finally, we give the
conclusion and discussion for the future prospect in the fourth section.

\section{Optical analogue of BTZ black holes}

The crucial point for the analogue of the black hole is the formation of the
horizons, which is usually dependent on the consideration of the black hole
spacetime as a moving fluid, i.e., the fluid as the background is flowing
along a direction to the region beyond the Newtonian escape velocity (i.e.,
the local velocity of sound), and the point where the velocity of background
fluid equals the sound velocity represents the horizon of the black hole. For
our purpose, the optical vortices are used since they can carry the angular
momentum. In order to simulate a rotating acoustic BTZ black hole, there
should be an effective method to form the inner horizon beside that realized
in the earlier simulation for the rotating black holes where the event horizon
and the ergosphere had been formed. In this section, we consider a
monochromatic laser beam propagates through a bulk medium, as in Ref.
\cite{fm08,vmf18}, and discuss how to relate it to a BTZ black hole metric.

We start with the nonlinear Schr\"{o}dinger equation (NLSE) in the paraxial
approximation, which governs the evolution of the electric field $E(x,y,z)$ of
the vortex beam,
\begin{equation}
\partial_{z}E=\frac{\mathrm{i}}{2k}\nabla_{\bot}^{2}E+\mathrm{i}\frac{kn_{2}%
}{n_{0}}E\left\vert E\right\vert ^{2}, \label{nlse}%
\end{equation}
where $z$ is the propagation direction, $n_{0}$ is the linear refractive index
and $n_{2}$ is the third-order nonlinear coefficient, $k=\frac{2\pi n_{0}%
}{\lambda}$ is the longitudinal wave number, $\left\vert E\right\vert ^{2}$ is
the optical intensity, and $\nabla_{\bot}^{2}E$ is defined with respect to
transverse coordinates $(x,y)$, which can accounts for diffraction while the
nonlinear term describes the self-defocusing effect.

To present the relation with the fluid dynamics, the electric field can be
written with the form, $E=\rho^{1/2}\mathrm{e}^{\mathrm{i}\phi}$. Thus, NLSE equation
(\ref{nlse}) becomes the hydrodynamic continuity and Euler equations
\cite{blv11,fm08},
\begin{equation}
\partial_{z}\rho+\nabla\cdot(\rho\mathbf{v})=0, \label{hc}%
\end{equation}

\begin{equation}
\partial_{z}\psi+\frac{1}{2}v^{2}+\frac{c^{2}n_{2}}{n_{0}^{3}}\rho-\frac
{c^{2}}{2k^{2}n_{0}^{2}}\frac{\nabla^{2}\rho^{1/2}}{\rho^{1/2}}=0, \label{ee}%
\end{equation}
where $c$ is the speed of light, the optical intensity $\rho$ corresponds to
the fluid density, $\mathbf{v}=\frac{c}{kn_{0}}\nabla\phi\equiv\nabla\psi$ is
the fluid velocity, and the third term is the quantum pressure which is
usually ignored in the linearized process for the derivation of the analogue
metric (see Eq. (\ref{obhm}) below), but it actually plays an important role
close to the horizons \cite{fs12}, determining the width of the horizon and
regularizing the otherwise singular solution.

We linearize the density and the velocity potential with $\rho=\rho_{0}+$
$\rho_{1}$, $\psi=\psi_{0}+\psi_{1}$, where $\rho_{0}$ and $\psi_{0}$ are
related to the background fluid determined by the equations of motion, and
$\rho_{1}$ and $\psi_{1}$ are small perturbations. Then, we put the new
velocity potential and mass density after perturbations into the Eqs.
(\ref{hc}) and (\ref{ee}). Finally, the two equations are combined into one
about the fluctuation $\psi_{1}$. It is found that the fluctuation $\psi_{1}$
satisfies a wave equation, $\nabla^{2}\psi_{1}=\left(  1/\sqrt{-g}\right)
\partial_{\mu}\left(  \sqrt{-g}g^{\mu\nu}\partial_{\nu}\psi_{1}\right)  =0$,
with the metric given as \cite{wgu81,fm08},
\begin{equation}
\mathrm{d}s^{2}=(\frac{\rho_{0}}{c_{s}})^{2}[-(c_{s}^{2}-v_{0}^{2}%
)\mathrm{d}t^{2}-2v_{r}\mathrm{d}r\mathrm{d}t-2v_{\theta}r\mathrm{d}%
\theta\mathrm{d}t+dr^{2}+(r\mathrm{d}\theta)^{2}],\label{obhm}%
\end{equation}
where $c_{s}$ is the local speed of sound and is related to the bulk pressure
which is derived from the nonlinear interaction. The terms $v_{r}$ and
$v_{\theta}$ represent the radial and tangential velocity components, with
their definitions as $v_{r}=\partial_{r}\psi_{0}$ and $v_{\theta}=\frac{1}%
{r}\partial_{\theta}\psi_{0}$, from which the total velocity is $v_{t}%
^{2}=v_{r}^{2}+v_{\theta}^{2}$. For the metric (\ref{obhm}), it can model a
spacetime about a black hole if the region $v_{r}>c_{s}$ exists where anything
cannot escape, and it can model a spacetime about the ergosurface of a
rotating black hole if the region $v_{t}>c_{s}>v_{r}$ exists where it is
impossible for an observer to remain stationary relative to a distant
observer. So, it is appropriate to simulate the properties of a rotating black
hole, as made in Ref. \cite{vmf18}, but the inner horizon can not be found in
their work. Here, we firstly see if the metric (\ref{obhm}) can match the BTZ
black hole metric in form.

The rotating BTZ black hole metric is given as \cite{btz92}%

\begin{align}
\mathrm{d}s^{2}  &  =-(-M+\frac{r^{2}}{l^{2}}+\frac{J^{2}}{4r^{2}}%
)\mathrm{d}t^{\prime}{}^{2}+\frac{\mathrm{d}r^{2}}{-M+\frac{r^{2}}{l^{2}%
}+\frac{J^{2}}{4r^{2}}}\nonumber\\
&  +r^{2}(\mathrm{d}\theta^{\prime}-\frac{J}{2r^{2}}\mathrm{d}t^{\prime})^{2},
\label{btzo}%
\end{align}
where $\Lambda=-\frac{1}{l^{2}}$ is the negative cosmological constant and the
units with $c=G=1$ are taken. $M$ and $J$ are the mass and angular momentum of
the black hole, and are given as%

\begin{equation}
M=\frac{r_{+}^{2}+r_{-}^{2}}{l^{2}},\qquad J=\frac{2r_{+}r_{-}}{l},
\label{mam}%
\end{equation}
where $r_{+}$, $r_{-}$ represent the positions of the outer and inner horizon
of the black hole, respectively. They are determined by $-M+\frac{r^{2}}%
{l^{2}}+\frac{J^{2}}{4r^{2}}=0$, which leads to the results, $r_{\pm}%
^{2}=\frac{Ml^{2}}{2}\left(  1\pm\sqrt{1-(\frac{J}{Ml})^{2}}\right)  $. We may
assume without loss of generality that $J\geq0$ and assume that $Ml\geq J$ to
ensure the existence of an event horizon at $r=r_{+}$. In particular, the
ergosurface is specified at the position $r_{E}=\sqrt{M}l=\sqrt{r_{+}%
^{2}+r_{-}^{2}}$.

To compare conveniently with the metric (\ref{obhm}), we make the
transformation for the BTZ black hole metric from the coordinates $t^{\prime
},\theta^{\prime}$ to $t,\theta$ according to the following forms,%

\begin{align}
\mathrm{d}t  &  =\mathrm{d}t^{\prime}-\frac{\sqrt{1-(-M+\frac{r^{2}}{l^{2}%
}+\frac{J^{2}}{4r^{2}})}}{-M+\frac{r^{2}}{l^{2}}+\frac{J^{2}}{4r^{2}}%
},\nonumber\\
\mathrm{d}\theta &  =\mathrm{d}\theta^{\prime}-\frac{J}{2r^{2}}\frac
{\sqrt{1-(-M+\frac{r^{2}}{l^{2}}+\frac{J^{2}}{4r^{2}})}}{-M+\frac{r^{2}}%
{l^{2}}+\frac{J^{2}}{4r^{2}}},
\end{align}

This changes the metric (\ref{btzo}) into the form,%

\begin{align}
\mathrm{d}s^{2}  &  =-(-M+\frac{r^{2}}{l^{2}})c_{s}^{2}\mathrm{d}t^{2}%
-2c_{s}\sqrt{1-(-M+\frac{r^{2}}{l^{2}}+\frac{J^{2}}{4r^{2}})}\mathrm{d}%
t\mathrm{d}r\nonumber\\
&  +Jc_{s}\mathrm{d}t\mathrm{d}\theta+\mathrm{d}r^{2}+r^{2}\mathrm{d}%
\theta^{2}, \label{btzn}%
\end{align}
where the speed of light is introduced again but with the local expression
$c_{s}$.

Thus, we can link the nonrelativistic metric (\ref{obhm}) with the BTZ black
hole metric (\ref{btzn}) up to the overall factor $(\frac{\rho_{0}}{c_{s}%
})^{2}$ by letting%

\begin{equation}
v_{r}=c_{s}\sqrt{1-(-M+\frac{r^{2}}{l^{2}}+\frac{J^{2}}{4r^{2}})}, \label{frv}%
\end{equation}

\begin{equation}
v_{\theta}=c_{s}\frac{J}{2r}. \label{fav}%
\end{equation}
It is seen that the metric (\ref{obhm}) is valid only in the range where the
radial velocity $v_{r}=c_{s}\sqrt{1-(-M+\frac{r^{2}}{l^{2}}+\frac{J^{2}%
}{4r^{2}})}$ is well defined, and this holds for $R_{-}<r_{-}<r_{+}<R_{+}$ with%

\begin{equation}
R_{\pm}^{2}=\frac{(1+M)l^{2}}{2}\left[  1\pm\sqrt{1-(\frac{J}{(1+M)l})^{2}%
}\right]  , \label{fbhb}%
\end{equation}
where $R_{\pm}$ represent the boundary that the analogue can be made. It is
easy to check for the consistency that the horizons which is determined in the
fluid aspect by the relation $v_{r}(r)=c_{s}(r)$ are just at $r_{\pm}$ that
determined in the BTZ black hole aspect. Similarly, it has the same result
with $v_{t}(r)=c_{s}(r)$ for the determination of the ergosurface. Moreover,
to ensure that the ergosphere can be observed in the analogue system, it
requires $r_{E}<R_{+}$, which leads to a new constrained condition
$Ml^{2}>J^{2}$. Therefore, under all these constrained conditions, it is
obtained that the acoustic metric realized by the optical vortex can be
regarded as the analogue of the BTZ black hole with the nearly perfect match
in form up to an overall factor.

In Fig.1, we present the actual conditions required for the fluid as the
realization of the analogue BTZ black hole, with the parameters $M=3$,
$J=1.3l$ ($<\sqrt{M}l$), which gives $r_{+}\approx1.69l$, $r_{-}\approx0.38l$,
$r_{E}\approx1.73l$, $R_{+}\approx1.98l$ and $R_{-}\approx0.33l$ by the
intersections of the line of the local speed of sound with the lines of
$v_{r}\left(  r\right)  $ and $v_{t}\left(  r\right)  $. Note that here
$c_{s}$ is assumed to be a constant without loss of generality, but in the
actual physical systems such as Bose-Einstein condensates \cite{gaz00} or the
optical vortex \cite{vmf18,fm08}, the velocity $c_{s}$ is not constant. In
fact, the critical point for the analogue is the realization of the relations,
i.e. $v_{r}(r)=c_{s}(r)$, $v_{t}(r)=c_{s}(r)$ as stated above. $R_{+}$ and
$R_{-}$ give the boundary beyond which the simulation of an acoustic rotating
BTZ black hole would be invalid. Obviously, they satisfy the constraint:
$R_{-}<r_{-}<r_{+}<r_{E}<R_{+}$. Besides the above mentioned constraints, we
still have to consider the continuity of the fluid flow in the practical
simulation. In the case here, it is required to fine tune the density as
$\rho\propto\frac{1}{r\sqrt{1-(-M+\frac{r^{2}}{l^{2}}+\frac{J^{2}}{4r^{2}})}}$
to satisfy the continuity equation $\nabla\cdot{(\rho\mathbf{v})}=0$.

\begin{figure}[ptb]
\centering
\includegraphics[width=1\columnwidth]{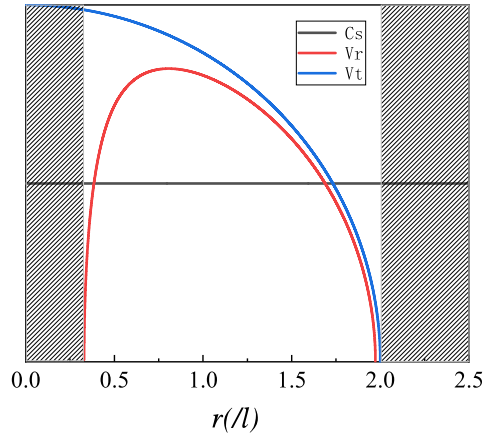} \caption{(Color online) The
velocities of the fluid as a function of the radius for the representation of
the rotating BTZ black hole analogue with $M=3$ and $J=1.3l$. The red line
represents the radial velocity which leads to the horizons at the
intersections with the black line of the sound speed. The blue line represents
the total velocity which gives the position for the ergosurface when it
matched with the sound speed. The shaded gray regions represent the invalid
range in which the acoustic analogue breaks down because the radial velocity
is ill defined. }%
\label{Fig1}%
\end{figure}

\section{Optical realization}

In the last section, we have presented how to simulate the BTZ black hole
using the fluid model. However, the required form for the radial and
tangential velocities is hard to be found in the real physical system. In the
earlier experiment \cite{vmf18}, a monochromatic laser beam, with the electric
field $E_{0}=\sqrt{\rho_{0}(r)}\mathrm{exp}(\mathrm{i}m\theta)$ where $m$ is
topological charge integer of the optical vortex, is used to simulate the
background spacetime as a photon fluid, in which the radial velocity
$v_{r}(r)=\frac{c}{kn_{0}}\frac{-\pi}{\sqrt{r_{0}r}}$ and tangential velocity
$v_{\theta}(r)=\frac{c}{kn_{0}}\frac{m}{r}$ is gotten respectively. This can
present the phenomena of rotating black holes like the ergosphere, but it
doesn't include the inner horizon which is also a vital component for the
structure of rotating black holes. In this section, we consider a special
Gauss optical vortex with the electric field
\begin{equation}
E_{0}=\sqrt{\rho_{0}(r)}\mathrm{exp}\left(  \mathrm{i}m\theta-2\mathrm{i}%
\pi\left(  Ar-B\mathrm{ln} r-Cr^{2}\right)  \right)  , \label{novf}%
\end{equation}
where $\rho_{0}(r)=\rho_{0}\left(  \frac{r}{\sigma}\right)  ^{1/2}%
\mathrm{exp}\left(  -\frac{2r^{2}}{\sigma^{2}}\right)  $ takes the
Gaussian-like form, and $\rho_{0}$, $A$, $B$, $C$ are the constants. The
corresponding phase is given as $\varphi(r)=-2\pi\left(  Ar-B\mathrm{ln}
r-Cr^{2}\right)  $, which leads to the new expressions for the radial and
tangential velocities,%
\begin{align}
v_{r}(r)  &  =\frac{-2\pi c}{kn_{0}}\left(  13-\frac{3}{r}-5r\right)
,\label{nfrv}\\
v_{\theta}(r)  &  =\frac{c}{kn_{0}}\frac{m}{r}, \label{nfav}%
\end{align}
when the optical vortex is injected into a self-defocusing medium (nonlinear
refractive index $n_{2}<0$) without cavity. Here, the radial velocity is
obtained by taking the radial derivative of the phase $\varphi$ and the
constants are taken as $A=13$, $B=3$, $C=2.5$ for matching the geometric
structure of the BTZ black holes in the last section, i.e. see the Fig. 2.

\begin{figure}[ptb]
\centering
\includegraphics[width=1\columnwidth]{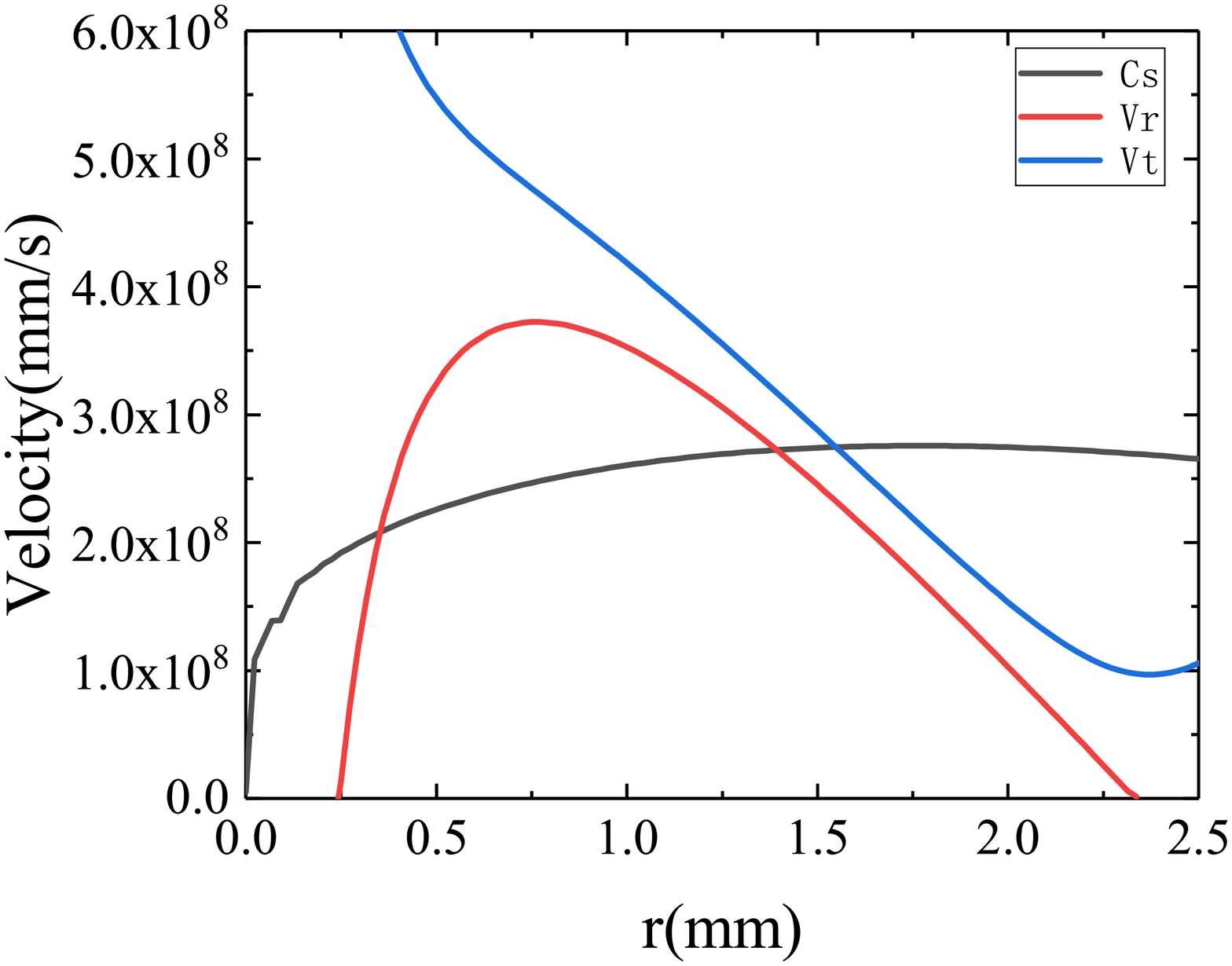}\newline%
\includegraphics[width=1\columnwidth]{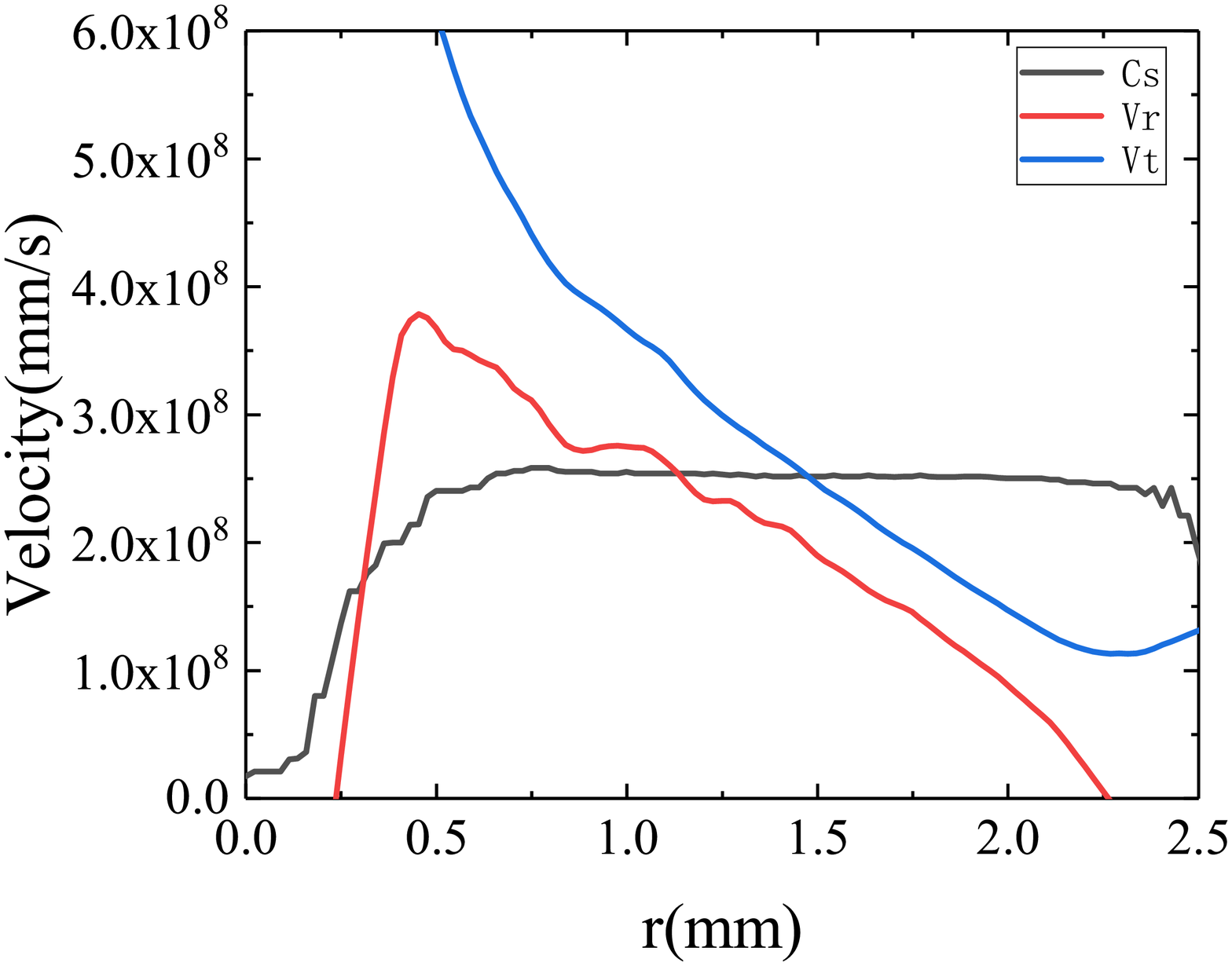}\newline\caption{(Color online)
The velocities of the fluid as a function of the radius obtained by
numerically calculating the evolution equation (\ref{nlse}) with the initial
velocities as in Eqs. (\ref{nfrv}) and (\ref{nfav}). The parameters are taken
according to the actual experimental requirements, i.e. the linear refractive
index $n_{0}=1.5$, the nonlinear coefficient $|\gamma|=4.4\times10^{-7}
cm^{2}/W$, the power of the laser $p=140 mW$, The beam waist of the laser
$\sigma=5 mm$, and the topological charge integer m=20. The upper (lower)
panel presents the results as the propagation distance of the optical vortex
into the medium is $0.5 mm$ ($14 cm$). The lines has the same meaning as in
Fig. 1. }%
\label{Fig2}%
\end{figure}

In Fig. 2, we present the numerical results with the evolution in Eq.
(\ref{nlse}) for the change of the radial velocity $v_{r}(r)$, the total
velocity $v_{t}\left(  r\right)  =\sqrt{v_{\theta}^{2}+v_{r}^{2}}$, and the
speed of sound $c_{s}(r)=\sqrt{c^{2}|\gamma|\rho_{0}(r)/n_{0}^{3}}$ (where
$\gamma$ is the nonlinear coefficient of the medium) with the radius $r$. It
is seen that the corresponding structure is determined by the relations, i.e.
$v_{r}(r)=c_{s}(r)$ for the horizons, and $v_{t}(r)=c_{s}(r)$ for the
ergosurface, which are seen by the intersections of velocity curves, i.e. the
left (right) intersection of velocity curves $v_{r}(r)$ and $c_{s}(r)$
determines the position of the inner (outer) horizon, and the intersection of
velocity curves $v_{t}(r)$ and $c_{s}(r)$ determines the position of the
ergosurface. This is nearly the same with the requirement of simulating BTZ
black holes as in Fig. 1, but the boundary of the valid analogue determined by
$v_{r}(r)=0$, or $R_{+}\approx2.34$ and $R_{-}\approx0.25$, is larger than
that for the analogue BTZ black hole in Fig. 1.

The upper panel of Fig. 2 presents the analogue structure at the distance of
$0.5$ millimeters for the propagation of the optical beam in the nonlinear
medium, which shows that the nonlinear interaction makes the analogue black
hole structure form quickly. When the beam propagates in the medium for a
distance of 14 centimeters, it is seen in the lower panel of Fig. 2 that the
analogue BTZ black hole structure is remained well. This shows that the
stability of analogue black hole structure for the whole process of the
optical vortex passing through the nonlinear medium, and it is advantageous
for the further studies about the analogous properties of black holes, i.e.
Hawking radiations. It is noted that the Hawking radiation had been discussed
before using the optical vortex and a very interesting phenomena about
resonance enhancement of Hawking radiation was found \cite{obf18} in which the
quantum potential was considered when the analysis is near the horizon.
Although they gave an attractive model including a white-hole horizon and the
corresponding ergosurface, and a black-hole horizon and the corresponding
ergosurface, the inner horizon required for the rotating BTZ black holes is
still not included. It is significant to see if this phenomena of Hawking
radiation enhancement exists for our suggested model which included a complete
BTZ black hole spacetime structure. This would be investigated in our future work.

Moreover, if the topological charge takes a smaller number, i.e. $m=2$, the
analogue BTZ black hole structure still exists, but the distance between the
outer horizon and the ergosurface is a little small and cannot be displayed in
the plot at the scale of millimeters. Anyway, we give an approximate fluid
model to present all the elements required by the BTZ black holes. The next
crucial question is: can the form in Eq. (\ref{novf}) about the optical vortex
be realized in the present experiment? In the following, we will interpret
this using the numerical simulation.

\begin{figure}[ptb]
\centering
\includegraphics[width=1\columnwidth]{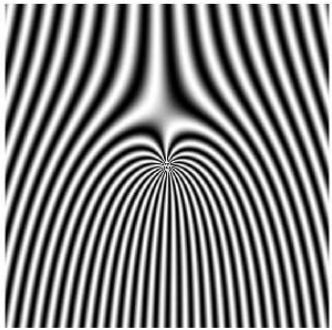} \caption{(Color online)
Hologram for the optical vortex satisfying the required phase given in Eq.
(\ref{novf}). The bifurcated part represents the topological charge integer
$m=20$. }%
\label{Fig3}%
\end{figure}

Due to the widespread technological applications in manipulating small
macroscopic particles \cite{hfr95}, the cooling and trapping of neutral atoms
\cite{krn98}, Bose-Einstein condensates \cite{wad00}, driving micro fabricated
gears for creating micrometre-scale motors \cite{fr01} and so on, there has
been increasing interest in the generation of optical vortex which has helical
wavefront and undefined phase at an isolated zero amplitude point in recent
years. Several methods for generating optical vortex have been reported,
including the use of computer generated holograms \cite{hmw92}, geometric mode
converters \cite{abw92,baw93}, spatial light modulators (SLM)
\cite{ckg02,cg03,wcz22}, and spiral phase plates \cite{vvk06,jfz09,cwz20}.
Here we suggest a method using liquid crystal spatial SLM to generate the
optical vortex with the requirement given in Eq. (\ref{novf}). The crystal
spatial SLM can regulate the amplitude and phase of the optical waves by the
electro-optical effect of the liquid crystal. In order to generate such an
optical vortex, ones have to obtain the hologram of the phase $\varphi(r)$ at
first, as given in Fig.3 using the numerical simulation, and then crave an
optical grating according to the hologram and load the grating into the
pure-phase SLM of liquid crystal. When a planar light is guided through the
SLM with the loaded holographic optical grating, the required optical vortices
would be generated by the exiting light. The phase and the near-field
intensity are simulated numerically, as given in Fig. 4 and Fig. 5. It is
noted that the center of optical vortex is dark in Fig. 5. However, when the
vortex beam is injected into the nonlinear self-defocusing medium, the central
dark core will shrink, and the bright ring will expand and become flatten for
the light intensity of the transverse profile, as seen in Fig. 2. In
particular, it will form optical solitons under certain experimental conditions.

\begin{figure}[ptb]
\centering
\includegraphics[width=1\columnwidth]{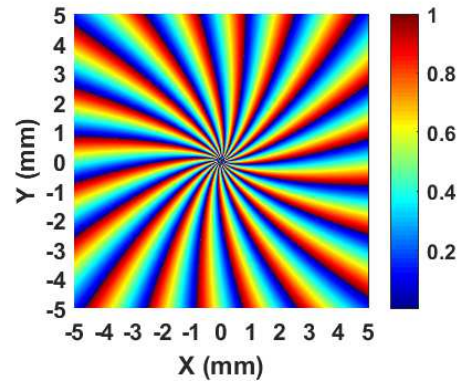} \caption{(Color online) The
simulated phase of the optical vortex. The phase increases according to the
color from blue to red. The repeated time of the color is the topological
charge integer $m=20$. }%
\label{Fig4}%
\end{figure}

\begin{figure}[ptb]
\centering
\includegraphics[width=1\columnwidth]{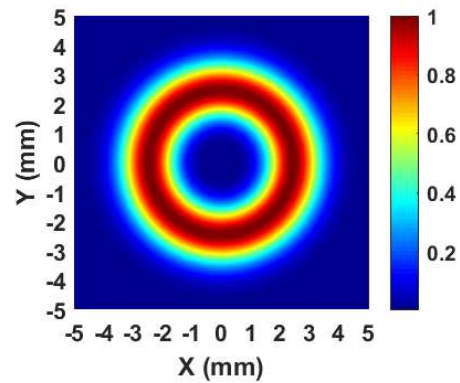} \caption{(Color online) The
simulated intensity of the optical vortex. The intensity increases according
to the color from blue to red.}%
\label{Fig5}%
\end{figure}

\section{Conclusion and Discussion}

In the paper, we review the earlier suggestion using the photon fluid in an
optical vortex to simulate the rotating black holes. It is noted that their
analogue did not include the inner horizon that is characteristic for the
rotating black hole. We use the same physical systems to give a kind of
analogue black hole including the inner horizon. At first, we compare the
acoustic metric obtained from the photon fluid model with the BTZ black hole
metric, and give the expressions for the flowing velocity of the fluid in the
radial and tangential direction. With these, we have described the spacetime
structure of the BTZ black hole in the analogue fluid model and determined the
conditions for the formation of the inner and outer horizon and the
ergosurface. However, the radial velocity in Eq. (\ref{frv}) is not integrable
along the radius and so it cannot find an exact expression for the phase of
the optical vortex. Fortunately, an approximate form for the phase has been
found, and all the structures required for the BTZ black hole can be
presented. We have also given a suggestion about how to generate the optical
vortex and obtained the properties about its wavefront and intensity using the
method of numerical simulation.

With regard to the interesting properties related to AdS/CFT or the string
theory that could be presented using the BTZ black hole, it is significant to
realized the analogue BTZ black hole in the lab which would provide a physical
platform for the corresponding exploration. As far as we know, there was a
suggestion to simulate the BTZ black hole in the real physical system, so our
work might make up a helpful suggestion for the future experiments about
analogue gravity.

\section{Acknowledgments}

We appreciate Prof. Daniele Faccio for his detailed and helpful replies on
their related work. This work is supported by the NSFC under Grant No. 11654001.

\end{document}